\documentclass{article}

\usepackage{arxiv}
\usepackage[table]{xcolor}
\usepackage[utf8]{inputenc} 
\usepackage[T1]{fontenc}    
\usepackage{hyperref}       
\usepackage{url}            
\usepackage{booktabs}       
\usepackage{amsfonts}       
\usepackage{nicefrac}       
\usepackage{microtype}      
\usepackage{lipsum}
\usepackage{graphicx}
\graphicspath{ {./images/} }

\title{SuperMeshing: A New Deep Learning Architecture for Increasing the Mesh Density of Metal Forming Stress Field  with Attention Mechanism and Perceptual Features}

\author{\textbf{Qingfeng Xu$^{1,3}$} \quad \textbf{Zhenguo Nie$^{1,2,}$}\thanks{Address all correspondences to these authors. (zhenguonie@tsinghua.edu.cn)} \quad \textbf{Handing Xu$^{1,4}$} \quad \textbf{Haosu Zhou$^{5}$}  \quad \textbf{Xinjun Liu$^{1,2,*}$}  \\ 

${^1}$The State Key Laboratory of Tribology \&\\ Tsinghua University (DME)-Siemens Joint Research Center for Advanced Robotics,\\ Department of Mechanical Engineering (DME), Tsinghua University, Beijing 100084, China\\
${^2}$Beijing Key Lab of Precision/Ultra-precision Manufacturing Equipments and Control,\\ Tsinghua University, Beijing 100084, China\\
${^3}$School of Computing and Information Systems. Melbourne School of Engineering,\\ University of Melbourne, Melbourne 3000, Australia\\
${^4}$School of Mechatronical Engineering, Beijing Institute of Technology, Beijing 100084, China\\
${^5}$Dyson School of Design Engineering, \\Imperial College London Exhibition Rd, London, SW7 2AZ, UK}
\begin{document}
\maketitle
\begin{abstract}
In stress field analysis, the finite element analysis is a crucial approach, in which the mesh-density has a significant impact on the results. High mesh density usually contributes authentic to simulation results but costs more computing resources, leading to curtailing efficiency during the design process. To eliminate this drawback, we propose a new data-driven mesh-density boost model named SuperMeshingNet that strengthens the advantages of finite element analysis (FEA) with low mesh-density as inputs to the deep learning model, which consisting of Res-UNet architecture, to acquire high-density stress field instantaneously, shortening computing time and cost automatically. Moreover, the attention mechanism and the perceptual features are utilized, enhancing the performance of SuperMeshingNet. Compared to the baseline that applied the linear interpolation method, SuperMeshingNet achieves a prominent reduction in the mean squared error (MSE) and mean absolute error (MAE) on test data, which contains prior unseen cases. Based on the data set of metal forming, the comparable experiments are proceeded to demonstrate the high quality and superior precision of the reconstructed results generated by our model.
The well-trained model can successfully show more excellent performance than the baseline and other methods on the multiple scaled mesh-density, including $2\times$, $4\times$, and $8\times$. With the refined result owning broaden scaling of mesh density and high precision, the FEA process can be accelerated with seldom cost on computation resources. We publicly share our work with full detail of implementation at \url{https://github.com/zhenguonie/2021_SuperMeshing_2D_Metal_Forming}.
\end{abstract}


\section{Introduction}
Numerical methods, for instance finite, element analysis (FEA), are widely employed in engineering domain\cite{fea_1,fea_2}. Compared with traditional experiments, the numerical simulation can optimize the design and process parameters at a low cost. In the process of FEA, especially the refined finite element mesh provides an essential mesh density for FEA to reach the precise value \cite{fea3,fea4}. 
However, to promote the accuracy of the simulation, higher mesh-density is required to be implemented during numerical simulation, which escalates the requirement of calculation resources, leading to the decrease of computational efficiency and increase of computing time.
Therefore, it is critical for FEA to generate the results that have high mesh-density while preserving on scaling down time consumption.

The prediction tasks for stress field analysis based on data-driven technology are inspired by contemporary achievement of deep learning for computer vision, natural language processing, and control \cite{imagenet2017krizhevsky, bert2018devlin, alphago2017}. Machine learning methods have been predominantly examined and investigated in multiple aspects of FEA. To predict steady flows in a representative, a capsule neural network was established in the area of fluid dynamics \cite{fea202kurtakoti}.
Besides, deep neural networks (DNN) also have been utilized to learn to the structural features to boost the performance in topology optimization \cite{tp2019oh,tp22019rawat,tp32020nie}. Furthermore, to estimate stress distribution, a fast deep learning approach is developed \cite{dlfea2018,stressgan2020jiang}. However, the competence of prediction works appear powerless ability of generalization and accuracy because of the enormous disparity of input and output data. On the other hand, to accelerate the computing productivity, deep learning models \cite{meshingnet, meshingnet2} are exploited to generate mesh automatically but without time saving during the process of FEA calculating.   
Therefore, against the experimental success of traditional meshing methods and machine learning methods, the following key challenges still persist. One of drawbacks is lacking of ability of generalization which measures the performance of the model to learn finite samples of a data distribution to reconstruct other samples. Secondly, the computing time is inevitable to be tremendous while aiming to generate a high quality outcome.

Inspired by the deep learning works on physical field prediction, a refined finite element simulation model based on deep neural network is developed, named SuperMeshingNet. SuperMeshingNet is trained with low mesh-density finite element analysis results as input which corresponding with the high mesh-density finite element analysis results as targets, while the same standard mesh-density of the finite element analysis results as outputs. The SuperMeshingNet learn the mapping relationship between the FEA result with low mesh-density and high mesh-density FEA result, which can be utilized to instantaneously generate various version of higher mesh-density FEA result. To boost our approach, multiple deep learning techniques have been manipulated on SuperMeshingNet, in where the ResUNet \cite{resnet, stressnet2018nie} constitutes the main structure, the attention mechanism \cite{attention_1,attention_2,attention_image} provides focusing of training, and the perceptual feature \cite{image_sr2015, image_sr2016, image_sr2017} promotes visual quality. 

With this model, a variety of high mesh density finite element meshes can be calculated by only consuming a microscopic amount of computing resources on generating lower mesh density finite element analysis results and model pre-training, which can be immediately give feedback to the designer, saving time. Meanwhile, by assessing on the test data set which consists of distinctive data compared with training set, the precision and the ability of generalization are validated. Moreover, the experiments examining SuperMeshingNet by competing performance with baseline and several models at diverse scaling factor, including $2\times$, $4\times$, and $8\times$. SuperMeshingNet achieves superior performance on all measure metrics, contrasted with other models in experiment.  
In addition, since deep learning network training is more dependent on matured learning infrastructures, the deep learning model migrates the calculation of finite element grids from traditional CPUs to cheaper GPUs, significantly reducing the computational cost.

\begin{figure}[!ht]
    \centering
    \includegraphics[width=0.5\linewidth]{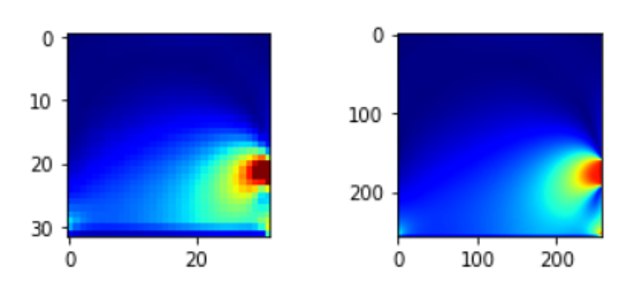}
    \caption{Low mesh density FEA result and High mesh density FEA result.}
    \label{fig:fea_lr_sr_introduction}
\end{figure}

Our main contributions are:
\begin{description}
  \item[$\bullet$] A new method that employing deep network models to generate high mesh-density simulation results through low mesh-density simulation results with seldom time-consuming.
    \item[$\bullet$] A new design of the output metrics measuring the quality of mapping relationship among various mesh-density scale of FEA results and evaluation matrices involving the geometric feature and perceptual feature of the FEA result.
  \item[$\bullet$] A hybrid neural network architecture with channel and spatial attention mechanism as the main structure of SuperMeshingNet that guaranteeing tremendous accuracy.
\end{description}
\section{Related Work}
Mesh density expands the information of a FEA result, similar with a image super-resolution work that deep learning based approaches have become the dominant position since the advance of convolution neural network for processing image. Meanwhile, attention mechanism that used in CNN achieves outstanding performance. This trend also shows in finite element analysis topic. Therefore, we review four parts of knowledge that related to our work: increasing mesh density, image based super-resolution, attention mechanism in deep learning and related knowledge about deep learning.
\subsection{Adaptive Mesh Refinement}
The meshing process in the finite element analysis process is an important issue. Some works \cite{automesh1999yokoyama,automesh2016cuilliere} have done a detailed study on how to quickly obtain a adaptive mesh density. Besides, several studies \cite{AMR_2, AMR_3, AMR_4, AMR_5} continued to propose various enhanced adaptive mesh refinement algorithms. But for designers, because these algorithms still requires multiple iterations to obtain higher numerical accuracy, decreasing in computing efficiency, there is no quick, simple, and intuitive way to determine the mesh density, and a large number of finite element calculations are still inevitable. 
\subsection{Image Based Super-resolution}
Supervised neural networks have become state-of-the-art work in super-resolution. The first CNN architecture to recover high-resolution(HR) images by low-resolution(LR) images was proposed by Dong \cite{image_sr2015}. Both image resolution increasing and FEA meshing density improving face a key problem: completion of missing information \cite{image_sr2015,image_sr2016,image_sr2017}. Therefore, drawing on the approach of image processing is of considerable significance to the information completion process in the finite element analysis process. The knowledge about image super-resolution, such as texture feature, also benefits further FEA process.

From the aspect of human perception, texture usually describes the quality of the HR images. To reconstruct reasonable texture detail, an edge-directed SR algorithm integrated with the strengths of the detail synthesis approach was proposed by Tai et al. \cite{texture_1}. Capturing redundancies of similar image groups in various scope by building a multi-scale dictionary shows the contribution on gaining texture detail \cite{texture_2}. Besides, Wang et al. employing spatial feature transform layers to generate more realistic and visually pleasing texture \cite{a_texture}.

\subsection{Attention Mechanism}
Commonly, to inform the model where to focus attention can be claimed as an instruction to allocate inclinable available computing and storage resources towards the most descriptive components of source data with \cite{attention_1}. Newly, the attention mechanism has been applied to deep neural networks in some excellent works, covering nature language processing \cite{attention_nlp} and understanding in images \cite{attention_image}. To rebuild the feature and attention maps, these works combined with a gating function \cite{attention_2}.

\begin{figure}[!ht]
    \centering
    \includegraphics[width=0.7\linewidth]{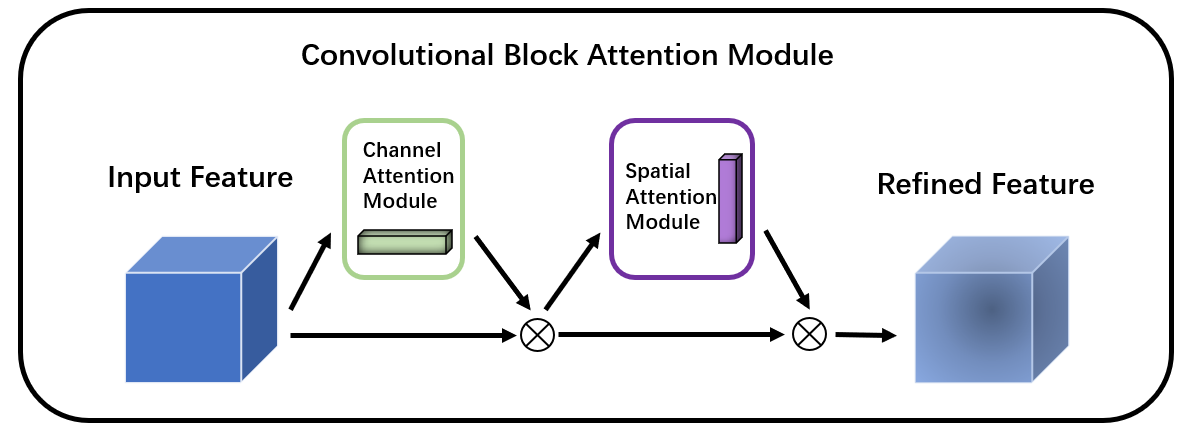}
    \caption{The overview of channel and spatial attention module in convolutional block \cite{tp32020nie}.}
    \label{fig:channel_spatial_attention}
\end{figure}

The representation of interests also is improved by the attention mechanism. Providing appropriate attention to different regions, spatial locations and channels can acquire better performance. SelNet \cite{attention_selnet} attaches a selection unit at the end of CNN layers. These units control the feature information that can be pass to next layer. Separately, the Residual Channel Attention Networks(RCAN) \cite{attention_rcan} focuses on channel attention. RCAN proposes RIR modules with a long skip connection to achieve channel attention which contains a similar structure with SqueezeNet.\cite{attention_squeezenet}. However, few techniques of attention mechanism have been developed to assistant FEA. 

\subsection{U-Net and ResNet}
\begin{figure}[!ht]
    \centering
    \includegraphics[width=0.3\linewidth]{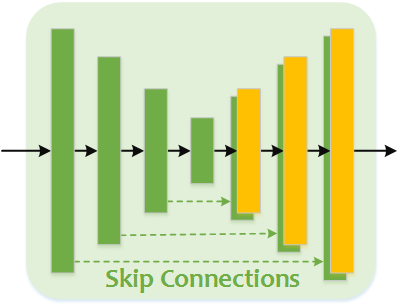}
    \caption{Architecture of the U-Net. The skip connections between networks are illustrated as dotted arrows \cite{stressnet2018nie}.}
    \label{fig:unet}
\end{figure}

 Figure \ref{fig:unet} shows the U-Net architecture\cite{unet2015long} which connect the down-sampling layers and up-sampling layers at various resolutions to deliver context information in preview layers. To predict the missing information in lower resolution, the skip-connection structure reflects the input data to extrapolated missing context, which is important to generator \cite{unet22015ronneberger}.

Deep network architectures have been show great contribution in performance, while it can be difficult to train. A powerful design strategy is the concept of residual blocks\cite{resnet2016he}.
The researchers of \cite{resnet2016he} claim that stacking layers by residual block should not damage the performance of network because the useless layers can be simplified with same performance.

\section{Technical Approach}

The input source data set is represented as ${x} \in \mathbb{R}^n$, and the target output data set ${y}\in\mathbb{R}^n$. The target of our research is to find the optimal weight ${w}$ in a neural network model ${F}$, which performs as a nonlinear regression function  ${F}({x},{w}) = {y}$, where ${x}$ donates the low mesh-density data and ${F}({x},{w})$ represent the procedure of reconstructing high mesh-density data. To acquired the optimal weight ${w}^*$, the distance between the target ${y}$ and the output of deep learning model ${F}({x},{w})$ is measured by MAE and minimized as demonstrated in Eq. (\ref{equ:general process}).

\begin{equation}\label{equ:general process}
{w}^{*} = {arg}{{min}_w}\frac{1}{n}\sum_{i=1}^{n}\mid{y}_i-{F}({x}_i,{w})\mid
\end{equation}

\begin{figure}[!ht]
    \centering
    \includegraphics[width=0.7\linewidth]{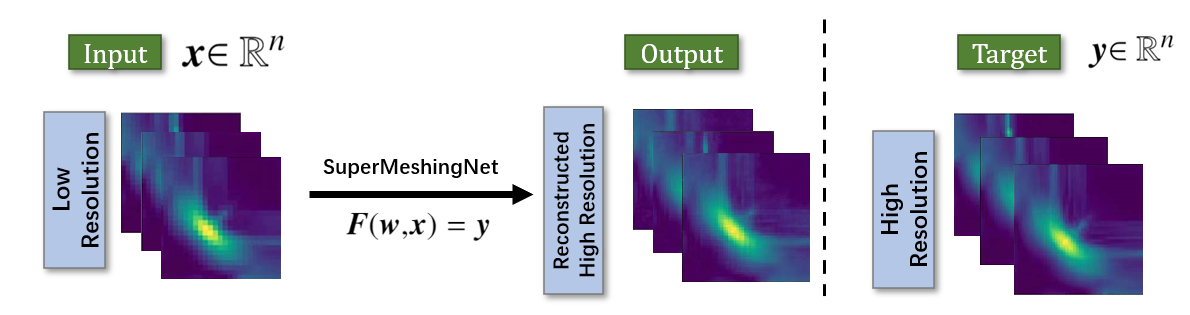}
    \caption{An overview of boosting mesh density analysis for FEA.}
    \label{fig:overview_process}
\end{figure}

The overall process of the mesh density increasing based on SuperMeshingNet is displayed in Figure \ref{fig:overview_process}. SuperMeshingNet is applied on the FEA, which the target data sets are computed based on finite element analysis theory. For the purpose of reconstructing the high mesh-density data, the low mesh-density data is utilized to train the SuperMeshingNet model to pursuit target data.

\subsection{Network Architecture}

\begin{figure}[!ht]
    \centering
    \includegraphics[width=0.8\linewidth]{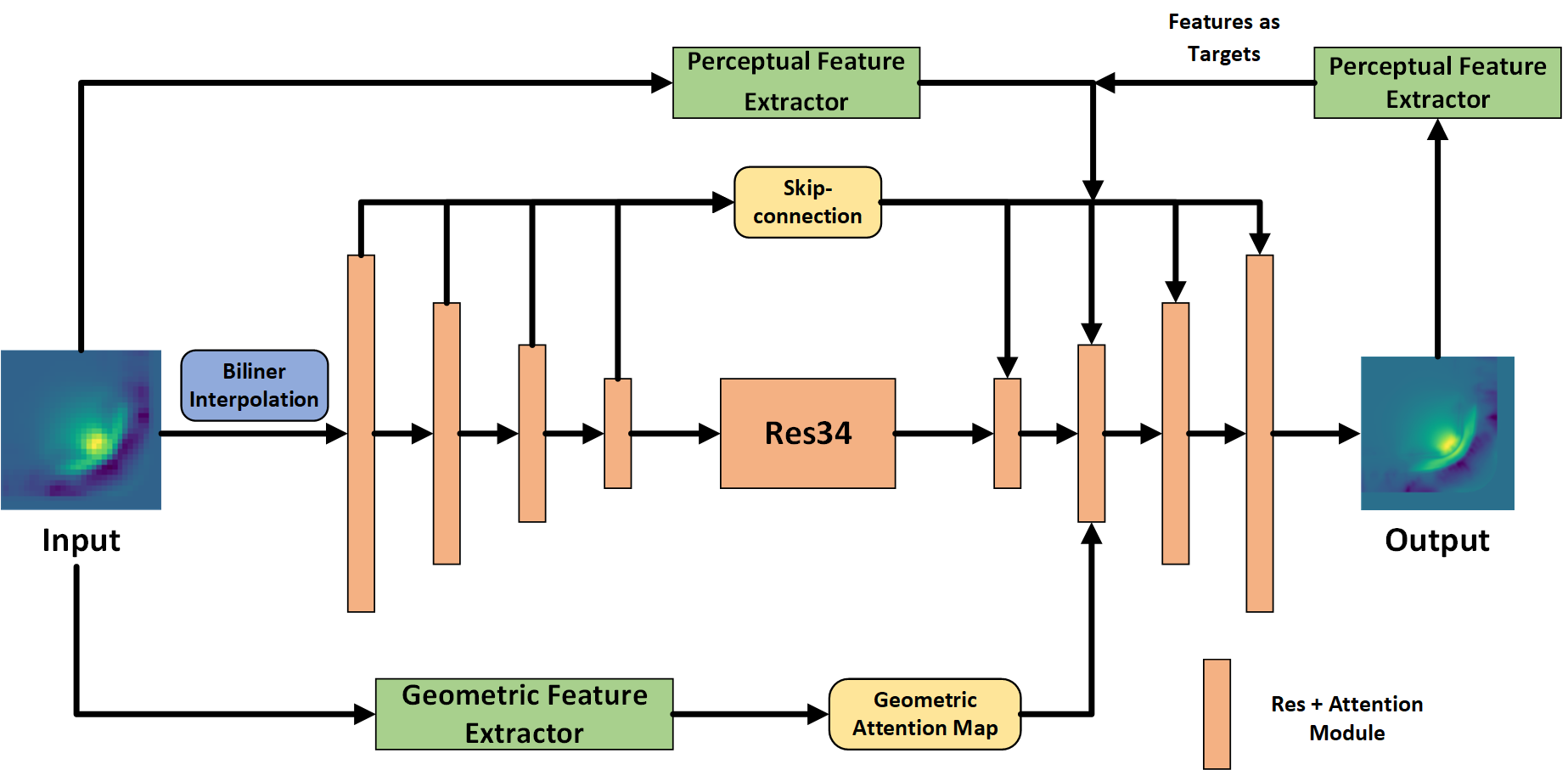}
    \caption{Architecture of the SuperMeshingNet.}
    \label{fig:architecture}
\end{figure}

The structure of the SuperMeshingNet is shown in Figure \ref{fig:architecture}. The main structure of the model is composed of ResNet with the channel and spatial attention modules, and the down-sampling layer and the up-sampling layer are connected by a structure that skip-connection. 

At the beginning of the model, the input is up-sampled to the same size as the output through a linear interpolation method, which can maintain the symmetry of the convolution process and contribute to achieving skip-connections. Another benefit of applying linear interpolation before convolution is avoiding another up-sampled process, optimizing the complexity of the model when dealing with the inputs that holding diverse mesh density. The data at the beginning of down-sampling maintains the same size with the ground truth, which decreasing to the half after a Res + Attention modules.  After four down-sampling Res + Attention modules, we added the Res34 structure to deepen the network and improve the effect \cite{resnet2016he}.

Up-sampling process appearances mirrored architecture of down-sampling process. Besides, a geometric attention map that contains the information of geometric feature generated from the training set is learned by a geometric extractor and added on the last third up-sampling module to highlight the geometric attention of the model on the image of the FEA result. Moreover, a perceptual feature extractor composed by ResNet is implemented to optimize the model by perceptual loss \cite{perceptual12016johnson, perceptual22019zhang, perceptual32016ledig} which enhancing the performance in feature space. To lighten the parameter of SuperMeshingNet and advance the training efficiency, two layers of down-sampling are utilized by geometric extractor and perceptual feature extractor.

\subsection{Attention Module}
\begin{figure}[!ht]
    \centering
    \includegraphics[width=0.7\linewidth]{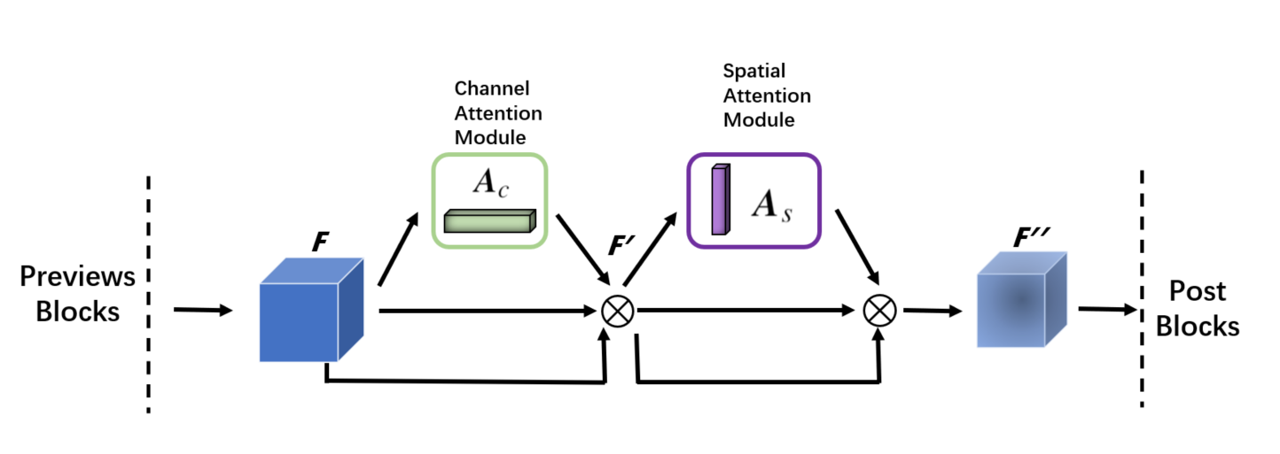}
    \caption{Implement of Res + Attention Module.}
    \label{fig:attention_module}
\end{figure}

$\mathbf{Channel\ and\ Spatial\ Attention\ Module.}$
The overall explanation of the attention module which constructed with ResBlock is shown in Figure \ref{fig:attention_module}. The feature map can be defined as ${F} \in \mathbb{R}^{C\times H\times W}$. Subsequently, the 1D channel attention map and the 2D spatial are characterized ${M_c} \in \mathbb{R}^{C\times 1\times 1}$ and ${M_c} \in \mathbb{R}^{1\times H\times W}$ respectively. 

\begin{equation}\label{equ:attention_11}
{F}'={M_c}\left({F}\right)\otimes{F}
\end{equation}
\begin{equation}\label{equ:attention_22}
{F}''={M_s}\left({F}'\right)\otimes{F}'
\end{equation}

\noindent where $\otimes$ is element-wise multiplication. As shown in Figure \ref{fig:attention_module}, the final clarified output $l{F''}$ comes from the input processing with channel attention value and spatial attention value multiplied. 

The channel attention module detects the meaningful features of a given input \cite{channelattention2018park}. The previous implementation also proves that the representation ability of networks can be promoted significantly by appealing both average-pooled and max-pooled features simultaneously \cite{cbam2018woo}. Therefore, we employ both technologies in channel attention:
\begin{equation}\label{equ:attention_1}
{M_c}\left({F}\right) = \sigma\left( W_1\left(W_0\left(F^c_{avg} \right)\right)+W_1\left(W_0\left(F^c_{max}\right)\right)\right)
\end{equation}

\noindent where the sigmoid function is represented as $\sigma$,  ${F^c_{max}}$ and ${F^c_{avg}}$ are the max-pooled features and average-pooled features. ${W_0}$ and ${W_1}$ share the weights for inputs followed by Leaky ReLU activation function. 

Dissimilar with channel attention, spatial attention pays more attention to the location of the effective informative part. After two pooling operations on feature map, there are two 2D maps: ${F^s_{avg}} \in \mathbb{R}^{1\times H\times W}$ and ${F^s_{max}} \in \mathbb{R}^{1\times H\times W}$. Subsequently, a standard convolution layer with $7\times7$ kernel size is utilized to process the concatenated inputs. The final refined output can be described as:
\begin{equation}\label{equ:attention_2}
{M_s}\left({F}\right) = \sigma\left( f^{7\times7}\left(\left[ {F^s_{avg}};{F^s_{max}}\right]\right)\right)
\end{equation}

\begin{figure}[!ht]
    \centering
    \includegraphics[width=0.3\linewidth]{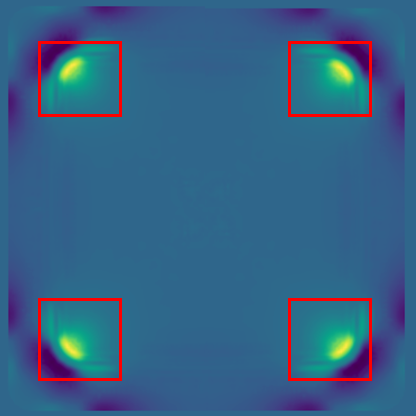}
    \caption{The geometric feature shown in FEA result of stamping.}
    \label{fig:geometric_feature}
\end{figure}

$\mathbf{Geometric\ Attention\ Map}$
At the same time, we discover that when finite element analysis is applied in the physical simulation process, the geometric shape is a salient feature that is easier to observe as shown in the Figure \ref{fig:geometric_feature}. Therefore, features related to geometry should be highlighted to assist the training and enhance the performance. 

To highlight the geometric feature, we fit high-dimensional features to form an attention map with ground truth geometric features. The feature map represents the performance of geometric attention on the training set, and in the reconstruction experiment after training, the attention map will no longer need to be optimized.

\subsection{Loss Function}
The loss function of our approach contains three parts, which can be interpreted as:

\begin{equation}\label{equ:total_loss}
\mathcal{L}_\mathit{overall}= \lambda_{c}\mathcal{L}_\mathit{c} + \lambda_{p}\mathcal{L}_\mathit{p} + \lambda_{g} \mathcal{L}_\mathit{g}
\end{equation}

$\mathbf{Content\ Loss.}$
The first and the most important part of overall loss is content loss which describe the quality of reconstruction. The $\mathcal{L}_1$  Loss \cite{l1loss2017janocha} is employed: 

\begin{equation}\label{equ:l1_loss}
\mathcal{L}_\mathit{c} = \mathit{l}\left(\mathit{x,y}\right)=\frac{1}{L\times W \times N}\sum_{i=1}^{N} \mid\mathit{x_i} - \mathit{y_i}\mid
\end{equation}
\noindent where ${L, W}$ donates the size of the data, ${N}$ means the batch size. ${x,y}$ represents the high mesh-density data generated by our model and ground truth, respectively. The $\mathcal{L}_1$  Loss has been indicated to be sharper for performance and more accessible for convergence compared to $\mathcal{L}_2$  Loss. Therefore, we choose $\mathcal{L}_1$  Loss to optimize the content performance.

$\mathbf{Perceptual\ Loss.}$
Some works \cite{perceptual12016johnson, perceptual22019zhang, perceptual32016ledig} in image super-resolution have adopted perceptual loss, which has been proved effective in improving visual quality. In these works, enhancing the similarity in feature space is the fundamental technology. We take advantage of perceptual loss, supplementing the loss function in feature space. The perceptual loss can be described as:
\begin{equation}\label{equ:perceptual_loss}
\mathcal{L}_\mathit{p}=\frac{1}{L_i\times W_i \times N} \mid\mathit{\phi_i^{NN}\left(x\right)} - \mathit{\phi_i^{NN}\left(y\right)}\mid^2
\end{equation}

\noindent where the ${L_i, W_i}$ is the shape of the feature map ${\phi_i^{NN}}$ in the $i$-th layer's of neural network (${NN}$) model, here the perceptual loss constraints the predicted high-resolution data to have similar detail to the ground truth in feature dimension, such as flatness and texture of the output. 

$\mathbf{Geometric\ Loss.}$
FEA is a simulation of the objective physical world, so its geometric features are salient features that can be observed by humans. We extract obvious geometric features by adding geometric attention to the model and apply them. In order to optimize the geometric attention map, we introduced geometric loss combine the Kullback-Leibler divergence \cite{kldivergence2020zhao} los: 

\begin{equation}\label{equ:geometric_loss}
\mathcal{L}_\mathit{g} =\frac{1}{L\times W \times N}\sum_{i=1}^{N} \mathit{y_i}\left(\log{y_i} - x_i\right)
\end{equation}

$\noindent$ where ${y_i}$ comes from the ground truth, ${x_i}$ is acquired by geometric extractor.


\section{Experiments}
The proposed SuperMeshingNet and the control group are trained on the same data set that contains aligned low mesh-density data and high mesh-density data which is also processed by linear interpolation to measure the performance of SuperMeshingNet.

\subsection{Data set}

\begin{figure}[!ht]
    \centering
    \includegraphics[width=0.6\linewidth]{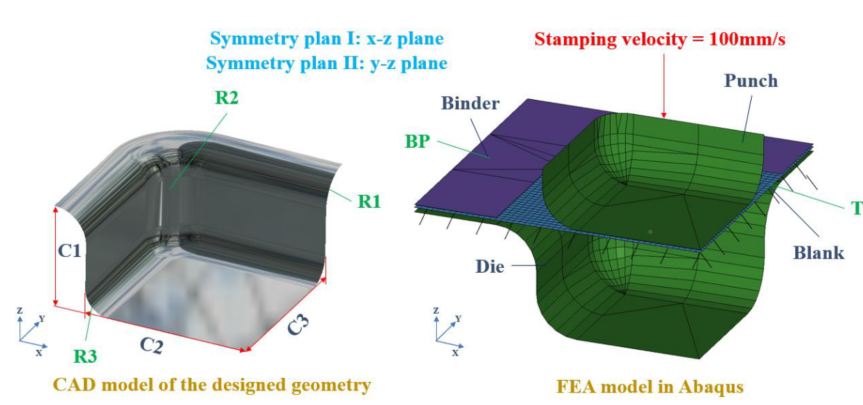}
    \caption{The CAD model and FEA model of a sample metal forming simulation \cite{zhou2020study}}
    \label{fig:sample model}
\end{figure}

In this study, the data set is comprised of stress fields in cold stamping process, which is generated by finite element method(FEM). As the symmetry of the simulation component, shown in Figure \ref{fig:geometric_feature}, \ref{fig:sample model}, presented data for the experiment is constructed by a quarter of the simulation result, and as-formed components expand into 2D-images. 

\subsection{Evaluation Metrics}

To evaluate the model, we use MAE shown in Eq.(\ref{equ:metric_mae}) and MSE shown in Eq.(\ref{equ:metric_mse}) for measuring the average and average squared divergence between the predicted high mesh-density data and the ground truth. 

\begin{equation}\label{equ:metric_mae}
\textrm{MAE} = \frac{1}{M} \sum_{i=1}^{M} \mid \sigma^{(i)} - \hat{\sigma}^{(i)} \mid 
\end{equation}

\begin{equation}\label{equ:metric_mse}
\textrm{MSE} = \frac{1}{M} \sum_{i=1}^{M} (\sigma^{(i)} - \hat{\sigma}^{(i)})^2
\end{equation}

\noindent where $M$ represents the number of test data. Meanwhile, $\sigma$ donates the real value of the stress field, and $\hat\sigma$ donates the predicted value of the stress field. Meanwhile, the distribution of the MAE and MSE of test cases are presented, which explains the performance at the level of the single case.

Moreover, some state-of-art processing results are exhibited to demonstrate the superior performance of the reconstruction process.

\section{Results and Discussions}
Our model is implemented by PyTorch, and trained on GPU (NVIDIA GeForce GTX 2080Ti). 
All the tested models select Adam \cite{adam2014kingma} for optimization. To convince the achievement of the SuperMeshingNet, three aspects of result is showed and discussed. Firstly, we present the performance of SuperMeshingNet that trained with the data set on various test cases to examine the effectiveness and generalization of our model. Subsequently, the performance of SuperMeshingNet, other neural networks working for super-resolution works, and linear interpolation methods are compared and discussed. Finally, aiming to explain the ability of computation resource-saving, the computing time under the same scenario is collected and presented.      

\subsection{Model Evaluation}
\begin{figure}[!ht]
    \centering
    \includegraphics[width=0.5\linewidth]{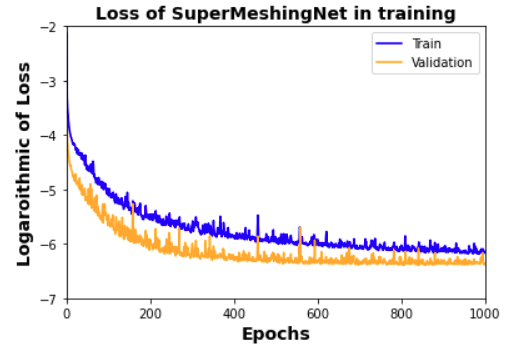}
    \caption{The loss of the SuperMeshingNet during training.}
    \label{fig:loss_in_training}
\end{figure}

As shown in Figure \ref{fig:loss_in_training}, the loss trend of the SuperMeshingNet in training, the validation loss restrains at 600 epochs, and train loss convergence after 800 epochs, which proves that the SuperMeshingNet performs reasonable convergence ability and indicates that the state-of-art training strategy should stop at 600 epochs to avoid over-fitting. 

\begin{figure}[!ht]
    \centering
    \includegraphics[width=0.7\linewidth]{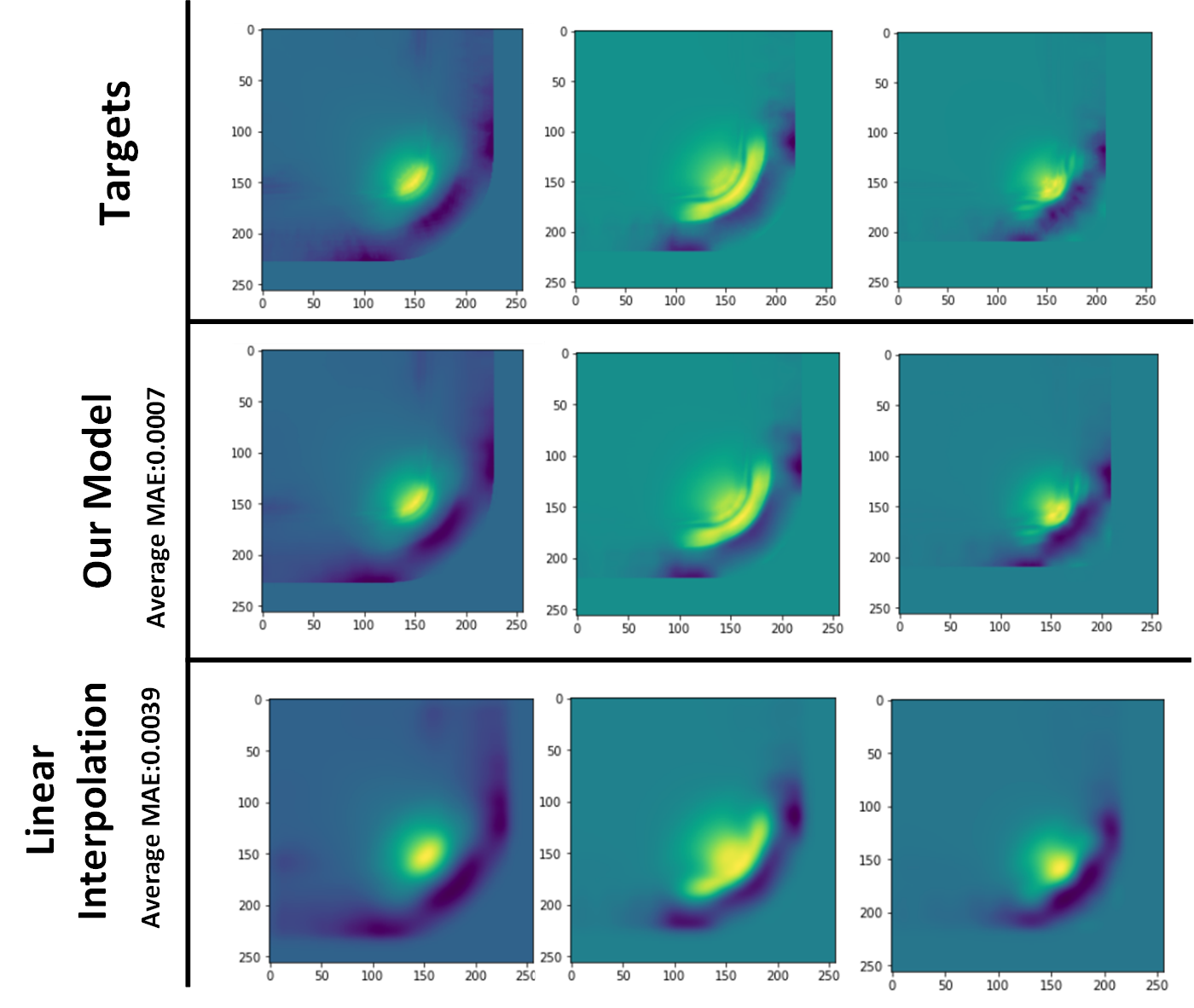}
    \caption{Some samples comparing result of our model with target and linear interpolation.}
    \label{fig:sample1}
\end{figure}
After completing the training of the model, we first observed its performance in the form of pictures on shape and texture and compared the results of the model reconstruction with the results obtained by the linear interpolation approach and the results obtained by the finite element calculation (ground truth). Figure \ref{fig:sample1} displays three cases from each method, where our model is SuperMeshingNet. Both our model and linear interpolation use the FEA result with the size of $32 \times 32$ as input to reconstruct an $8\times$ larger contemporary image which shape is $256 \times 256$. From the picture, it is not difficult to find that although the linear interpolation method can retain part of the shape information in the target data, the overall performance has been completely roughened due to the inaccuracy of details and textures, and the result is quite blurred, which is obviously not able to employ further application. On the contrary, our model's results of reconstruction are very consistent with the ground truth in terms of texture and shape, as well as the distribution of stress values. Compared with the linear interpolation average MAE error, the analytical product of our model is six times as low as, reaching 0.0007. For the finite element analysis process, our analytical results are similar to FEA calculations, which can predict the distribution trend of the force, and the error is minimal, so from the user's point of view, it will not affect the prediction of the force field trend.
\begin{figure}[!ht]
    \centering
    \includegraphics[width=0.7\linewidth]{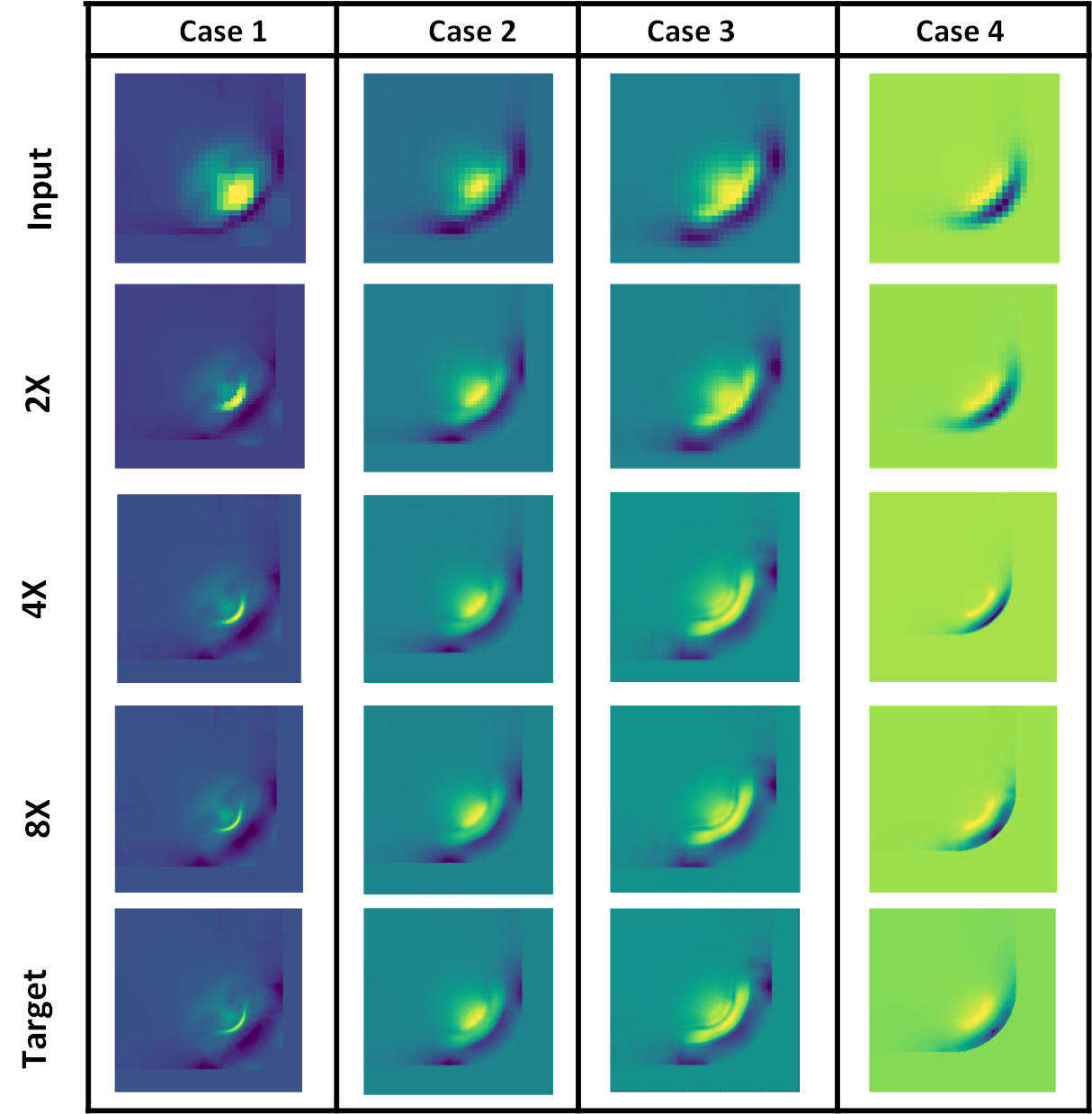}
    \caption{The sample of different degradation methods.}
    \label{fig:sample}
\end{figure}

More visual comparison is presented in Figure \ref{fig:sample}, where various examples reconstructed by our model to manifest the high quality of the outcomes. These examples include the $32\times32$ size to 2 times, 4 times, and 8 times mesh density of growth process. 
The results suggest that in the process of improving the mesh density of the input data, our model can match the actual value with their respective reconstruction methods. Therefore, the final 8-magnification reconstruction results based on all methods perfectly fit the real values in terms of overall details.
\subsection{Accuracy and Performance}
In this section, we verify the effectiveness of the SuperMeshingNet on different scaling factors by comparing five techniques, including baseline calculated by linear interpolation, ResNet, ResNet with skip-connection (ResUNet), ResUNet with attention mechanism, and SuperMeshingNet. In addition, the accuracy is presented by comparing various models. The SuperMeshingNet contains mainly four parts: skip-connection, attention module, perception feature extractor for feature synthesis, channel and spatial attention module, and the geometric feature attention map. The ablation of the experiment also can be suggested in Table \ref{tab:resolution_level}. After the skip-connection is employed, the MAE dropped to $4.758\times10^{-4}$ from $6.495\times10^{-4}$ when the scaling factor is 2. When channel and spatial attention module is utilized, the MAE can be decreased to $4.158\times10^{-4}$, which verifies the effectiveness of the attention mechanism. Subsequently, when the feature synthesis and geometric feature attention are progressively added, perceptual features will be enhanced, amplifying the performance to $3.203\times10^{-4}$. Furthermore, the same conclusion can be summarized from the scaling factor of $4\times$ and $8\times$.

\begin{table}[!ht]
    \caption{The performance of different mesh density.}
    \centering
    \begin{tabular}{c ccc}
        \toprule
        Model &$2\times$&$4\times$&$8\times$ \\
        
        \hline
        Baseline &$7.694\times 10^{-3}$&$3.581\times 10^{-3}$&$1.275\times 10^{-3}$\\
        ResNet & $9.714\times 10^{-4}$ &$7.923\times 10^{-4}$ &$6.495\times 10^{-4}$\\
        Res+$U$ &$7.770\times 10^{-4}$ &$6.174\times 10^{-4}$ &$4.758\times 10^{-4}$\\
        Res+$U$+$A$ &$\cellcolor{blue!40}5.957\times 10^{-4}$&$\cellcolor{blue!40}5.412\times 10^{-4}$&$\cellcolor{blue!40}4.158\times 10^{-4}$\\
        SMNet &$\cellcolor{yellow!40}5.465\times 10^{-4}$&$\cellcolor{yellow!40}4.682\times 10^{-4}$&$\cellcolor{yellow!40}3.203\times 10^{-4}$\\
       
        \hline
        \multicolumn{4}{p{8.5cm}}{\raggedright Note: 2$\times$, 4$\times$, and 8$\times$ are the scaling factors, and the metrics used is MAE. Baseline represents the linear interpolation, SMNet is the SuperMeshingNet, $U$ is the skip-connection structure, $A$ is the attention module.}\\
        \bottomrule
      \end{tabular}
     \label{tab:resolution_level}
\end{table}

\begin{figure}[!ht]
    \centering
    \includegraphics[width=0.6\linewidth]{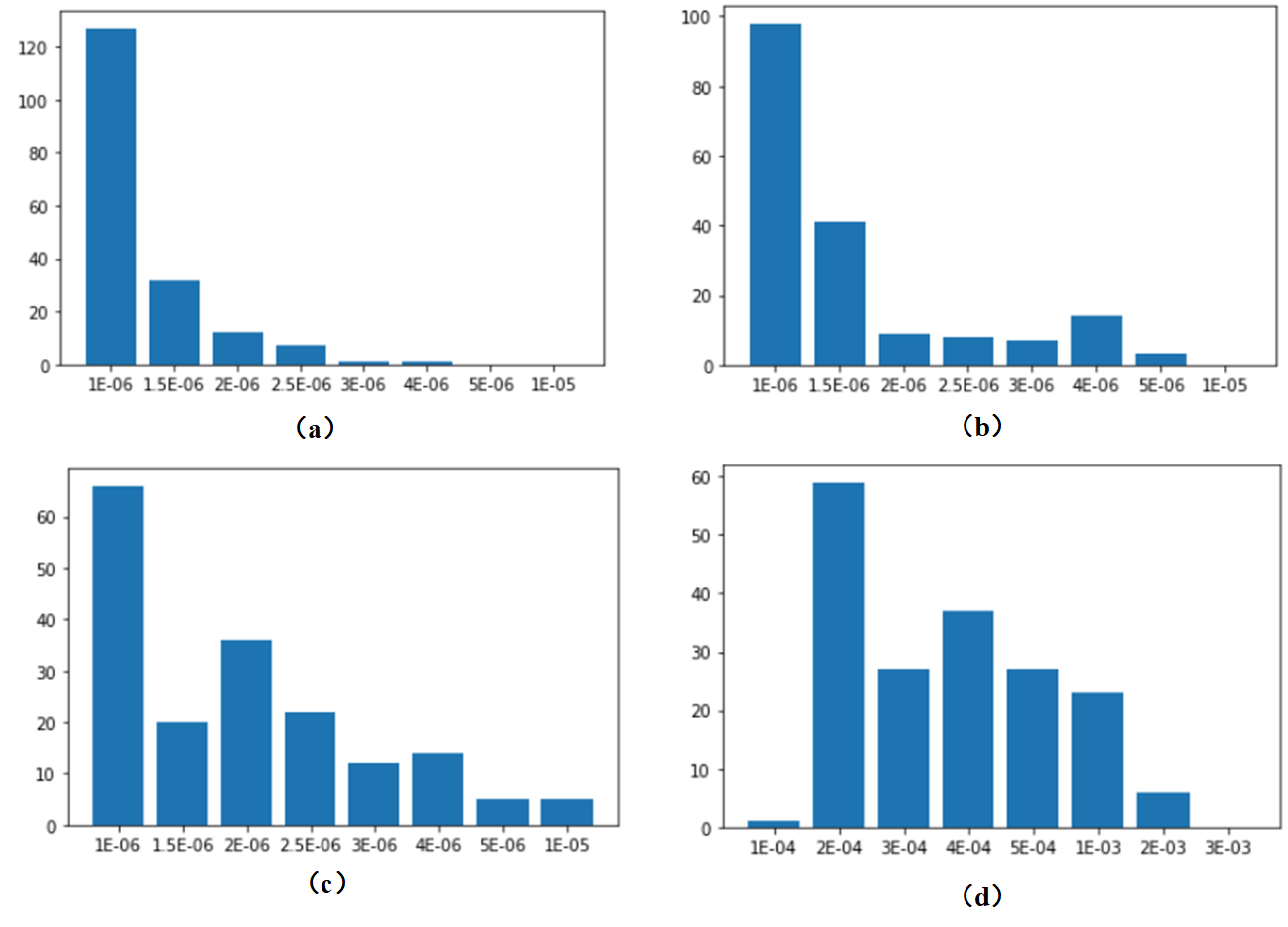}
    \caption{The distribution of the MAE of the four models, ${a}$: SuperMeshingNet, ${b}$: ResUNet+Attention Mechanism, ${c}$ :ResNet, and ${d}$: linear interpolation.}
    \label{fig:distribution}
\end{figure}

Figure \ref{fig:distribution} illustrates the distribution of the MSE after figuring out all test cases. Contrasting the cases acquired by linear interpolation in Figure \ref{fig:distribution} $\left({d}\right)$, which most cases range in $10^{-4} $ to $ 3\times10^{-3}$, the deep learning models constrain the MSE lower than $1\times10^{-5}$, proving that the deep learning methods performs more outstanding outcome compared with traditional linear interpolation. Furthermore, from the Figure \ref{fig:distribution} $\left({a}\right)$,  except few exceptions, the SuperMeshingNet restricts the MES of three quarters test set to be lower than  $10^{-6}$, while other cases tend to be lower than ResUNet with attention mechanism in the Figure \ref{fig:distribution} $\left({b}\right)$ and ResNet in the Figure \ref{fig:distribution} $\left({b}\right)$, convincing the superior generalization of the SuperMeshingNet.

The complementary justification is demonstrated in Table \ref{tab:mae_mse}, which shows both MAE and MSE of five techniques on training and test set. All experiments in Table \ref{tab:mae_mse} are performed with a scaling factor of $2\times$ low mesh-density and high mesh-density data pair. The yellow cells donate the lowest error while blue cells donates the second lowest error. As displayed in the comparison results, the SuperMeshingNet significantly outperforms both state-of-the-art MAE and MSE on training and test data set. Among the methods which aim the achieve better information loss, the skip-connection architecture and attention module contributes to reestablish the FEA data in higher quality, proving by the decreasing of the MAE and MSE. The MAE and MSE comparison results demonstrate the superiority of our proposed SuperMehsingNet over the other approaches.
\begin{table*}[!ht]
    \caption{Comparison and selection of architecture reconstruct result.}
    \centering
    \begin{tabular}{cl cc cc}
        \toprule
        \multicolumn{2}{c}{Metrics} & \multicolumn{2}{c}{MAE} & \multicolumn{2}{c}{MSE}\\
        \cmidrule(r){1-2} \cmidrule(r){3-4} \cmidrule(r){5-6}
        \textnumero & Model & Training  & Test & Training & Test\\
        \midrule
        
        1 & Baseline &$3.504\times 10^{-3}$&$3.526\times 10^{-3}$&$1.753\times 10^{-5}$&$1.783\times 10^{-5}$\\
        2 & ResNet &$5.963\times 10^{-4}$&$6.495\times 10^{-4}$&$1.054\times 10^{-6}$&$1.325\times 10^{-6}$\\
        3 & Res+$U$&$4.571\times 10^{-4}$ &$4.758\times 10^{-4}$ &$1.157\times 10^{-6}$&$1.274\times 10^{-6}$\\
        4 & Res+$U$+$A$&\cellcolor{blue!40}$4.196\times 10^{-4}$ &\cellcolor{blue!40}$4.152\times 10^{-4}$ &\cellcolor{blue!40}$8.605\times 10^{-7}$&\cellcolor{blue!40}$8.885\times 10^{-7}$\\
        5 & SMNet  &\cellcolor{yellow!40}$3.145\times 10^{-4}$ &\cellcolor{yellow!40}$3.203\times 10^{-4}$ &\cellcolor{yellow!40}$4.495\times 10^{-7}$&\cellcolor{yellow!40}$5.138\times 10^{-7}$\\
       
        \midrule
        \multicolumn{6}{p{11.5cm}}{\raggedright Note: Baseline represents the linear interpolation, SMNet is the SuperMeshingNet, $U$ is the skip-connection structure, $A$ is the attention module.}\\
        \bottomrule
      \end{tabular}
      \label{tab:mae_mse}
\end{table*}

\subsection{Computation Allocation} 
Another superior character of our model is time-saving. For each FEA calculation which the mesh density is $256\times256$ in Ansys Workbench, $T_{F256}$ spends 15.21 seconds on CPU (R7-4800HS), but only 0.41 seconds is cost by FEA computation with the size of $32\times32$, $T_{F32}$. For the SuperMeshingNet, the time spent on network training is relatively long, and the time spent in this experiment $T_{train}$ is $7,962.61$ seconds. With the trained model, the time required to reconstruct the low mesh-density data to high mesh-density is extremely small, $T_{r}$, only 0.012 seconds is spent. The $T_S$ and $T_F$ are the total time that the SuperMeshingNet and FEA need, which can be claimed as: 
\begin{equation}\label{equ:supermeshingnet_time}
{T_S} = T_{train} + N \times T_r +N\times T_{F32}
\end{equation}
\begin{equation}\label{equ:fea_time}
{T_F} = N\times T_{F256}
\end{equation}

\noindent where ${N}$ donates the number of data to be processed, which is named workload. In addition, because the SuperMeshingNet breaks the limitations of standard finite element calculations on the CPU and migrates it to the GPU-based deep learning framework for calculation. Therefore, the cost required for calculation is reduced because not only of time but also the reduction in equipment prices. 
\begin{equation}\label{equ:cost}
{Cost} = Total\ Time \times Price\ per\ second 
\end{equation}
\begin{table}[!ht]
    \centering
    \caption{The time and cost that spend on different workload}
    \label{tab:time_cost}
\begin{center}
\begin{tabular}{ |c|c c|c c|c| } 
\hline
\multicolumn{1}{c}{} & \multicolumn{2}{c}{SuperMeshingNet}& \multicolumn{2}{c}{FEA}\\
\hline
Workload & Time(s) & Cost(\$) & Time(s) & Cost(\$)\\
\hline
1 & 7,962.63 & 0.99 & 15.21 & 0.002 \\ 
$10^2$ & 8,004.81 & 1.00 & $1.52 \times10^3$  & 0.2 \\ 
$10^4$ & $1.22 \times10^4$ & 1.52 & $1.52 \times10^5$ & 20\\ 
$10^6$ & $4.29 \times10^5$ & 64.35 & $1.52 \times10^7$ & 2,000\\
\hline
\end{tabular}
\end{center}
\end{table}
Table \ref{tab:time_cost} shows the performance of the time and price when the two calculation methods refer to the CPU service price of $\$0.56$ per hour and GPU service of $\$0.49$ per hour, which is identical to the performance of our experimental equipment provided by Google Cloud. As can be seen from Table \ref{tab:time_cost}, since the time allocation on training the model, the total time and overhead of the SuperMeshingNet are relatively high when there are fewer processing tasks. However, with the continuous increase of processing tasks, it can clearly see that the time and cost of traditional finite element calculation are at a considerable disadvantage. Therefore, using the low mesh-density finite element analysis result under the condition of ensuring absolute accuracy and then reconstructing it to high mesh-density using the SuperMeshingNet can significantly reduce the time and cost of the entire finite element analysis process.
\subsection{Limitations and Future Work}
The SuperMeshingNet has excellent performance in automatically improving the density of finite element analysis. This technology can significantly reduce the calculation time and cost while maintaining a low MAE and extreme deviation. However, future works still can be extended. 


The Transformer \cite{transformer2017vaswani} structure with attention as the main structure has developed rapidly, and important applications have been born in various fields, including the super-resolution field \cite{transformersr2020yang,transformersr22019kasem}. The geometric attention structure used in this article draws on some of the ideas of transformer. In future work, the complete transformer structure can be utilized to build the model and promote model innovation. In addition, though SuperMeshingNet accomplish the superior performance on 2D data, solving the problem of 3D model which consuming more computing resources is valuable as a future direction. To reconstruct 3D data, it is appropriate to replace the network architecture by graph neural network (GNN), which is applicable to 3D data.

\section{Conclusions}
We present a new deep learning model named SuperMeshingNet to reconstruct the FEA outcomes with low mesh-density to the high mesh-density results where the meshing and computation cause immense cost. With a trained model and low mesh density data, the high mesh-density results can be produced immediately with high accuracy, contributing to the efficiency of FEA in the stress field. 
SuperMeshingNet draws self-attention module and perceptual features, also employs deepened ResNet and skip-connection to optimize the model and boost performance. 
The model integrated these techniques are proven to be constructive to outperform other frameworks in generalization and accuracy by comparative experiments. Extensive experiments demonstrate the superiority of our work when compromising scaling factors of $2\times$, $4\times$, and $8\times$. Concurrently, compared with the traditional finite element calculation and other works, our model can effectively save computing time and cost. Embedding SuperMeshingNet into FEA, our approach can ensure expeditious computational process and marvelous spatial precision simultaneously under multiple scaling factors. 


\bibliographystyle{unsrt}



\end{document}